\documentclass[useAMS,usenatbib]{mn2e}

\usepackage{graphicx} 
\usepackage{txfonts}
\usepackage{lscape}
\usepackage{mathrsfs}
\usepackage{nicefrac}

\title{On the Initial Mass Function and tilt of the Fundamental Plane \\ of massive early-type galaxies}

\author[C. Grillo and R. Gobat]{C. Grillo$^{1,2}$\thanks{cgrillo@usm.lmu.de} and R. Gobat$^{3}$\\
$^{1}$Max-Planck-Institut f\"ur extraterrestrische Physik, Giessenbachstr., D-85748 Garching bei M\"unchen, Germany\\
$^{2}$Universit\"ats-Sternwarte M\"unchen, Scheinerstr. 1, D-81679 M\"unchen, Germany\\
$^{3}$CEA Saclay, DSM/DAPNIA/Service d'Astrophysique, F-91191 Gif-sur-Yvette Cedex, France}

\begin{document}

\date{Accepted --. Received --; in original form --}

\pagerange{\pageref{firstpage}--\pageref{lastpage}} \pubyear{--}

\maketitle

\label{firstpage}

\begin{abstract}
We investigate the most plausible stellar Initial Mass Function (IMF) and the main origin of the tilt of the Fundamental Plane (FP) for old, massive early-type galaxies. We consider a sample of 13 bright galaxies of the Coma cluster and combine our results with those obtained from a sample of 57 lens galaxies in the same luminous mass range. We estimate the luminous mass and stellar mass-to-light ratio values of the sample galaxies by fitting their SDSS multi-band photometry with composite stellar population models computed with different dust-free, solar-metallicity templates and IMFs. We compare these measurements and those derived from two-component orbit-based dynamical modelling. The photometric and dynamical luminous mass estimates of the galaxies in our sample are consistent, within the errors, if a Salpeter IMF is adopted. On the contrary, with a Kroupa or Chabrier IMF the two luminous mass diagnostics differ at a more than 4$\sigma$ level. For the massive Coma galaxies, their stellar mass-to-light ratio scales with luminous mass as the corresponding effective quantities are observed to scale on the FP. This indicates that the tilt of the FP is primarily caused by stellar population properties. We conclude that old, massive lens and non-lens early-type galaxies obey the same luminous and dynamical scaling relations, favour a Salpeter IMF, and suggest a stellar population origin for the tilt of the FP. The validity of these results for samples of early-type galaxies with different age and mass properties still remains to be tested.
\end{abstract}


\section{Introduction}

It is now commonly accepted that the most massive ($M_{*} > 10^{11}\,M_{\odot}$) galaxies in the local Universe are mainly ellipticals, composed of uniform old stellar populations that are completely formed at $z \sim 1$. For these objects, a possible formation picture consists in a very rapid ``monolithic'' collapse (e.g., \citealt{lar74}) followed by some additional moderate star formation and merging activity, as observed in ellipticals from $z \sim 0$ to $z \sim 1$ (e.g., \citealt{con07}). Although mass seems to drive for the most part the galaxy star formation history (SFH, e.g., \citealt{cow96}), environmental effects have also observable consequences (e.g., \citealt{gob08}). Understanding the initial formation and the subsequent evolution of massive galaxies is thus an important test for the $\Lambda$CDM paradigm.

Strong gravitational lensing is an invaluable tool to study the internal structure (e.g., \citealt{rus05}) and the luminous and dark matter composition (e.g., \citealt{gri08b,gri09b}) of massive early-type galaxies. Several studies (e.g., \citealt{gri09}, hereafter G09a; \citealt{tre06}) have proved that early-type lens galaxies are an unbiased sub-sample of massive early-type galaxies. Hence, massive non-lens as well as lens early-type galaxies can be used to conclude on their physical properties, like the stellar IMF, and the origin of the observed scaling relations. In particular, the FP (\citealt{djo87}; \citealt{dre87}), a scaling law between effective radius, central stellar velocity dispersion, and average surface brightness, and its interpretation (known as the tilt of the FP) in terms of a systematic increase of the galaxy effective mass-to-light ratio with effective mass have been extensively studied (e.g., \citealt{cio96}). However, for old, massive early-type galaxies, it is still not clear which IMF is the most suitable one (e.g., \citealt{van08} and references therein) and if the tilt of the FP is mainly determined by variations in the galaxy stellar population features (e.g., \citealt{all09}), dark matter content (e.g., \citealt{tor09}), or structural properties (e.g., \citealt{ber02}).

This letter is organised as follows. In Sect. 2, we describe how and why we select a sample of massive early-type galaxies of the Coma cluster. In Sect. 3, we compare photometric and dynamical modelling to measure the amount of luminous mass in the galaxies and analyse the consequences of adopting different IMFs. In Sect.~4, we explore for the galaxies in the sample the relation between stellar mass-to-light ratio and luminous mass and discuss the implications of this relation in the tilt of the FP. In Sect. 5, we combine the results on massive non-lens (Coma) early-type galaxies with those obtained in previous works on massive lens (SLACS) early-type galaxies. In Sect. 6, we summarise our analysis. In this study, a concordance cosmology ($H_{0}=70$ km s$^{-1}$ Mpc$^{-1}$, $\Omega_{\mathrm{m}}=0.3$, $\Omega_{\Lambda}=0.7$) is assumed and all logarithms have base 10 and adimensional arguments, obtained by dividing the dimensional quantities by the corresponding measurement units.

\section{The sample}

We concentrate on a sample of 13 bright early-type galaxies of the Coma cluster. The sample is composed of flattened, rotating and non-rotating, elliptical and lenticular galaxies for which detailed axisymmetric, orbit-based, dynamical modelling of kinematic and photometric observations was performed by Thomas et al. (\citeyear{tho07,tho09}, hereafter T07 and T09). Among the galaxies studied in T09, we select those that have multi-wavelength (\emph{ugriz} bands) photometric measurements from the SDSS\footnote{http://www.sdss.org/}.

The choice of this sample is dictated by several different reasons. First, we aim at comparing published dynamical and new photometric luminous mass estimates to check the reliability of the two methods and to investigate on the IMF of these galaxies. Second, we propose to relate two samples of non-lens (Coma) and lens (SLACS\footnote{http://www.slacs.org/}, see \citealt{bol06}) early-type galaxies with relatively similar physical characteristics.

\begin{table}
\centering
\caption{Dynamical and photometric luminous mass and stellar mass-to-light ratio estimates for a sample of 13 early-type galaxies of the Coma cluster.}
\label{tab1}
\begin{scriptsize}
\begin{tabular}{ccccc} 
\hline\hline \noalign{\smallskip}
NGC/IC & $\mathrm{log}(M_{*,\mathrm{dyn}})$ & $\mathrm{log}(M_{*}^{\mathrm{Sal,M}})$ & $\mathrm{log}(M_{*}^{\mathrm{Kro,M}})$ & $\mathrm{log}(M_{*,0}^{\mathrm{Sal,M}}L_{B,0}^{-1})$ \\
\noalign{\smallskip} \hline \noalign{\smallskip}
4952 & $11.60 \pm 0.12$ & $11.34^{+0.04}_{-0.06}$ & $11.18^{+0.03}_{-0.06}$ & $0.85^{+0.06}_{-0.04}$ \\
4944 & $11.13 \pm 0.12$ & $11.18^{+0.17}_{-0.06}$ & $11.04^{+0.19}_{-0.09}$ & $0.55^{+0.17}_{-0.06}$ \\
4931 & $10.73 \pm 0.13$ & $10.89^{+0.06}_{-0.04}$ & $10.72^{+0.05}_{-0.06}$ & $0.55^{+0.07}_{-0.04}$ \\
4926 & $11.58 \pm 0.12$ & $11.60^{+0.03}_{-0.05}$ & $11.43^{+0.02}_{-0.05}$ & $0.95^{+0.03}_{-0.05}$ \\
4908 & $11.43 \pm 0.12$ & $11.34^{+0.04}_{-0.06}$ & $11.18^{+0.03}_{-0.06}$ & $0.89^{+0.04}_{-0.06}$ \\
4045 & $11.23 \pm 0.12$ & $11.04^{+0.04}_{-0.05}$ & $10.87^{+0.04}_{-0.04}$ & $0.89^{+0.04}_{-0.05}$ \\
4860 & $11.56 \pm 0.13$ & $11.41^{+0.03}_{-0.03}$ & $11.26^{+0.02}_{-0.05}$ & $0.92^{+0.03}_{-0.04}$ \\
3947 & $10.81 \pm 0.13$ & $10.54^{+0.02}_{-0.16}$ & $10.36^{+0.02}_{-0.16}$ & $0.82^{+0.02}_{-0.16}$ \\
4839 & $12.06 \pm 0.14$ & $11.72^{+0.03}_{-0.03}$ & $11.54^{+0.01}_{-0.03}$ & $0.97^{+0.03}_{-0.04}$ \\
4827 & $11.59 \pm 0.12$ & $11.49^{+0.04}_{-0.04}$ & $11.32^{+0.04}_{-0.04}$ & $0.92^{+0.03}_{-0.04}$ \\
4807 & $11.04 \pm 0.12$ & $11.20^{+0.05}_{-0.06}$ & $11.04^{+0.04}_{-0.06}$ & $0.82^{+0.06}_{-0.06}$ \\
4871 & $11.02 \pm 0.15$ & $10.90^{+0.05}_{-0.05}$ & $10.74^{+0.04}_{-0.06}$ & $0.86^{+0.05}_{-0.05}$ \\
4841A & $11.69 \pm 0.16$ & $11.51^{+0.04}_{-0.04}$ & $11.34^{+0.04}_{-0.04}$ & $0.90^{+0.04}_{-0.04}$ \\

\noalign{\smallskip} \hline
\end{tabular}
\end{scriptsize}
\begin{list}{}{}
\item[References --]T09.
\end{list}
\end{table}

\section{The Initial Mass Function}

For the early-type galaxies in the sample, axisymmetric two-component dynamical models computed by superposition of orbits (\citealt{sch79}) were developed in T07 and T09 to reconstruct the galaxy luminous and dark matter composition. First, assuming three different inclination angles, each observed 2D luminosity profile was deprojected to derive a 3D luminosity profile and convert it into a luminous mass density distribution through a constant stellar mass-to-light ratio ($\Upsilon_{*}$). Then, a total mass density distribution was calculated by adding to the luminous component a dark matter one, which was parametrised as a NFW (\citealt{nav96}) or a logarithmic (\citealt{bin81}) profile. Finally, the best-fitting model was determined following the prescriptions of the maximum entropy technique (\citealt{ric88}) in superposing orbits in the gravitational potential associated to the total mass density distribution (for further details, see T07 and references therein). The galaxy luminous mass values $M_{*,\mathrm{dyn}}$, that are shown here in Table \ref{tab1}, were measured from the multiplication of the $R$-band luminosities and the best-fitting $\Upsilon_{*}$, where the minimum-$\chi^{2}$ value of the stellar mass-to-light ratio was chosen between the two different families of dark matter haloes.

We measure the same physical quantity (luminous mass) by fitting the galaxy SEDs with composite stellar population models. The SEDs are retrieved from the SDSS archive and are composed of 5 bands extending from 354 to 913 nm. The extinction-corrected total magnitudes (modelMag) correspond to magnitudes measured through equivalent apertures in all the bands and thus provide unbiased galaxy colours, in the absence of colour gradients. We use Maraston's (\citeyear{mar05}; indexed M) and Bruzual \& Charlot's (\citeyear{bru03}; indexed BC) dust-free templates at solar metallicity and adopt Salpeter (\citeyear{sal55}; indexed Sal), Kroupa (\citeyear{kro01}; indexed Kro), and Chabrier (\citeyear{cha03}, indexed Cha) IMFs. We assume a delayed exponential star formation history (SFH) and perform the SED fits on a three-parameter (age, star-formation time-scale, and luminous mass) grid of models (for further details, see G09a). The best-fitting luminous mass values for Maraston's templates and Salpeter ($M_{*}^{\mathrm{Sal,M}}$) and Kroupa ($M_{*}^{\mathrm{Kro,M}}$) IMFs are shown in Table~\ref{tab1}. 

\begin{figure}
  \centering
  \includegraphics[width=0.46\textwidth]{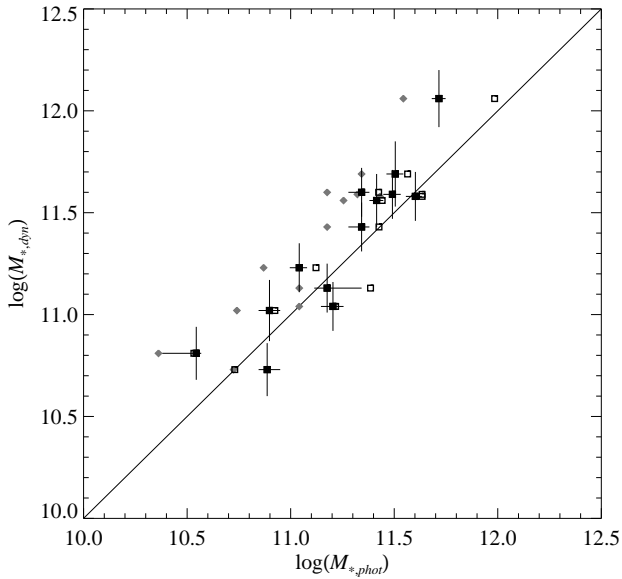}
  \caption{Comparison of the luminous mass estimates determined from two-component dynamical modelling ($M_{*,\mathrm{dyn}}$) and SED fitting ($M_{*,\mathrm{phot}}$) for a sample of 13 early-type galaxies of the Coma cluster. The values measured using Maraston's dust-free templates at solar metallicity and assuming a Salpeter and a Kroupa IMF are represented, respectively, with squares and diamonds. The empty squares represent the Salpeter photometric luminous mass values after a simple luminosity correction performed with the HyperLeda database. The 1$\sigma$ error bars and the one-to-one correlation line are shown.}
  \label{fi01}
\end{figure}

In Fig. \ref{fi01}, we plot for the 13 galaxies of our sample the dynamical and photometric luminous mass estimates reported in Table \ref{tab1}. In Table \ref{tab2}, we show the average value of the dynamical over photometric luminous mass ratio $\langle M_{*,\mathrm{dyn}} / M_{*,\mathrm{phot}} \rangle$ obtained with all the different templates and IMFs. We observe from Fig. \ref{fi01} and Tables \ref{tab1} and \ref{tab2} that the models with a Kroupa and Chabrier IMF give luminous mass measurements that are significantly smaller than the dynamical ones. In particular, the average ratio of the latter over the former deviates from one by more than 5$\sigma$. On the contrary, the models with a Salpeter IMF provide luminous mass estimates that are in general consistent with those determined from dynamics. Despite the overall good agreement of these two luminous mass diagnostics when a Salpeter IMF is adopted, we note that for some galaxies in the sample the dynamical mass values are moderately larger than the photometric ones. 

Although the SDSS photometry is very accurate, the large spatial extension of the light distribution of the galaxies in the sample and the assumption of a de Vaucouleurs profile to estimate the total magnitudes may require significant corrections in the photometric quantities (e.g., \citealt{kor09}). By comparing our best-fitting values of the $B$-band luminosity with those taken from the public database HyperLeda\footnote{http://leda.univ-lyon1.fr/} (\citealt{pat03}), we observe that not all the light belonging to the galaxies may have been included in the SDSS photometry. In the simplified hypothesis that the fraction of light missed in the five SDSS bands is constant and equal to that estimated from this comparison in the $B$-band, we can correct our photometric luminous mass estimates. This results in the empty squares plotted in Fig.~\ref{fi01} and in average values of $M_{*,\mathrm{dyn}} / M_{*,\mathrm{phot}}$ of $1.14 \pm 0.10$ and $1.67 \pm 0.15$ for Salpeter and Kroupa IMFs, respectively, adopting Maraston's templates.

A reason why the luminous mass measurements obtained through dynamical modelling can be slightly overestimated is the fact that a small fraction of mass of the dark matter component may have been associated to the luminous one. In fact, for a given set of observations of a galaxy, if the density distribution of its dark matter component resembles closely that of its luminous component, then the dynamical models may fit equally well by overestimating the luminous mass term at the expense of the dark matter one. Related to the previous point, the assumption on the parametrisation of the dark matter density distribution have a small, but not negligible, influence on the luminous mass estimates. For instance, in T07 and T09 the dynamical models with a logarithmic dark matter halo provide best-fitting values of the stellar mass-to-light ratio that are always equal or larger than those with a NFW halo and the majority of the dynamical luminous mass of the sample galaxies were derived from the logarithmic best-fitting $\Upsilon_{*}$. On the opposite side, possible underestimates of the photometric luminous mass determinations may be related to dust extinction or metallicity values lower than the solar one in the galaxies of the sample. These two effects would result in lower IR fluxes that in turn would give lower mass measurements. Several studies (e.g., \citealt{ret06}) have however tested the validity of the dust-free and solar-metallicity approximation. We remark that the discussed limitations in dynamical and SED modelling may justify the differences observed by comparing dynamical and photometric mass estimates with a Salpeter IMF, but not the much larger discrepancies obtained with a Kroupa or Chabrier IMF.

\begin{figure}
  \centering
  \includegraphics[width=0.46\textwidth]{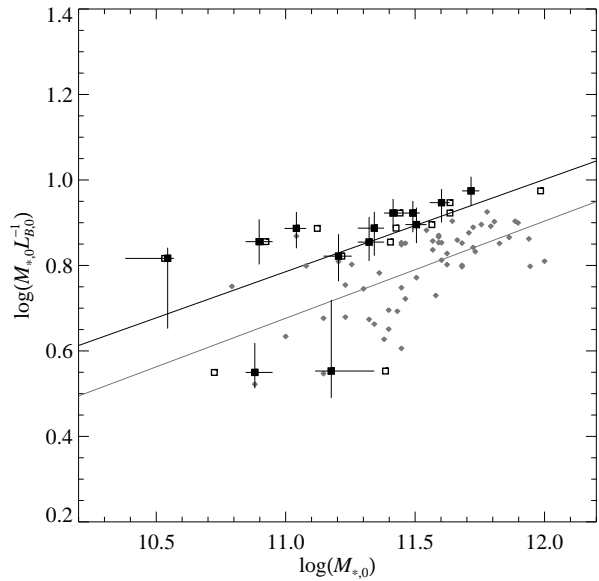}
  \caption{The $B$-band stellar mass-to-light ratio versus the luminous mass estimates at redshift $z=0$ for the early-type galaxies of the Coma (squares) and SLACS (diamonds) samples. The values and their 1$\sigma$ error bars are measured by using Maraston's dust-free templates at solar metallicity and assuming a Salpeter IMF. The empty squares represent the photometric luminous mass values of the Coma galaxies after a simple luminosity correction performed with the HyperLeda database. The best-fitting correlation lines for the two samples are shown.}
  \label{fi02}
\end{figure}

\begin{table*}
\centering
\caption{Important quantities derived from different composite stellar population models and IMFs.}
\label{tab2}
\begin{tabular}{ccccc} 
\hline\hline \noalign{\smallskip}            
 & Sal,M & Sal,BC & Kro,M & Cha,BC \\
\noalign{\smallskip} \hline \noalign{\smallskip}
$\langle M_{*,\mathrm{dyn}} / M_{*,\mathrm{phot}} \rangle$ & $1.34 \pm 0.13$ & $1.43 \pm 0.16$ & $1.96 \pm 0.19$ & $2.33 \pm 0.24$ \\
$\varrho [\mathrm{log}(M_{*,0}),\mathrm{log}(M_{*,0}L_{B,0}^{-1})]$ & 0.53 (0.06) & 0.54 (0.05) & 0.53 (0.06) & 0.57 (0.04) \\
$\varsigma [M_{*,0},M_{*,0}L_{B,0}^{-1}]$ & 0.72 ($<$0.01) & 0.66 ($<$0.01) & 0.69 ($<$0.01) & 0.55 ($<$0.01) \\
$\delta$ & $0.22 \pm 0.10$ & $0.16 \pm 0.07$ & $0.21 \pm 0.10$ & $0.18 \pm 0.08$ \\
\noalign{\smallskip} \hline
\end{tabular}
\begin{list}{}{}
\item[Notes --]For the the Pearson linear $\varrho$ and Kendall rank $\varsigma$ correlation coefficients, we show their values and, in parentheses, the probabilities that an equal number of measurements of two uncorrelated variables would give values of these coefficients higher than the measured ones.
\end{list}
\end{table*}

\section{The tilt of the Fundamental Plane}

The time evolution of the best-fitting SFHs (given by the parameters of age and star-formation time-scale of the best-fitting models for the SEDs) allows us to measure for all the sample galaxies their luminous mass ($M_{*,0}$) and stellar mass-to-light ratio in the $B$-band ($M_{*,0}L_{B,0}^{-1}$) at redshift $z = 0$. Since the Coma galaxies are local galaxies, the values of these two quantities do not differ significantly from those measured at the actual galaxy redshifts. In Table \ref{tab1} and in Fig. \ref{fi02}, we show the luminous mass and stellar mass-to-light ratio values that we obtain with Maraston's templates and a Salpeter IMF.

Although the points are relatively scattered, we observe in Fig. \ref{fi02} that the two quantities are correlated: very massive galaxies display larger stellar mass-to-light ratios than less massive galaxies. The values reported in Table \ref{tab2} of the Pearson linear ($\varrho$) and Kendall rank ($\varsigma$) correlation coefficients (for definitions, see \citealt{sal06}) prove that the luminous mass and the stellar mass-to-light ratio are statistically correlated at more than a 99\% confidence level (linearly at a 95\% confidence level). We note that the presence of a larger scatter at the low mass end is connected to differences in the galaxy stellar ages, as highlighted in T09. In passing, we mention that even if less evident the same correlation is found between the luminous mass and stellar mass-to-light ratio values determined in T09 from dynamical modelling. If we fit the scaling relation $M_{*,0}L_{B,0}^{-1} \propto M_{*,0}^{\delta}$, as done in Fig. \ref{fi02}, we obtain that our photometric measurements give a value of the exponent $\delta$ of $0.22 \pm 0.10$ (see Table \ref{tab2}). As reported in Table \ref{tab2} and shown in Fig. \ref{fi02}, we remark that this last result is not sensitive to the adopted IMF and only minimally influenced by the luminosity correction that we have performed with the HyperLeda data and discussed in the previous section. 

Based on $B$-band measurements presented in \citet{jor96} and on the FP analysis performed by \citet{ben98} in a much larger sample of Coma galaxies, \citet{tre06} finds that the effective (dynamical) mass-to-light ratio increases with the effective (dynamical) mass raised to the power of $0.21 \pm 0.02$. From directly observed quantities in a sample of nearly 9000 local early-type galaxies of the SDSS, \citet{ber03} also concludes that the dynamical mass-to-light ratio scales as the dynamical mass to the power of $0.22 \pm 0.05$. This relation is often used to parametrise the tilt of the FP. The extremely good agreement on the scaling relations between stellar and dynamical mass-to-light ratios and masses indicates that stellar population effects are the primary source of the tilt of the FP. This conclusion is also supported by the study of \citet{all09} on the FP of Coma galaxies.

By looking at the values determined by T09 of the fraction of dark matter over total mass and total mass included within a sphere of radius equal to the effective radius of our galaxies, we do not find a statistically significant correlation. Hence, an increment of the dark matter fraction going from massive to very massive galaxies does not seem to play an important role in defining the tilt of the FP. Just a secondary role may also be associated to a deviation from homology in the structural properties of early-type galaxies, as recently demonstrated by two studies on different samples of the SAURON (\citealt{cap06}) and SLACS (\citealt{bol08}) surveys. 

\section{Coma and SLACS galaxies}

The galaxies of the Coma sample studied here result to be comparable to those of the SLACS sample analysed in G09a in terms of luminous mass properties and as far as the main origin of the tilt of the FP is concerned. 

In agreement with the results of \citet{gri08} and G09a obtained from two samples of luminous and red lens early-type galaxies with luminous mass values in the same range of our Coma galaxies, we find that the choice of Maraston's or Bruzual \& Charlot's templates does not change remarkably the photometric mass measurements, as long as the same IMF is adopted. This means that systematic differences between these two models due to, for example, their different treatment of the late stages of stellar evolution do not play a significant role for these samples. Moreover, our photometric luminous mass measurements based on the accurate SDSS multi-colour photometry are consistent, within the errors, with those obtained from Schwarzschild and stellar dynamics plus lensing modelling only if a Salpeter IMF is chosen. We mention that for old galaxies (like those in our samples) a Salpeter IMF and the ``bottom-light'' IMF proposed by \citet{van08} are in good accord on the values of the stellar mass-to-light ratio.

After correcting for passive evolution, 15 lens galaxies of the SLACS survey were proved to occupy the same FP as non-lens galaxies of the Coma cluster (\citealt{tre06}). For a bigger galaxy sample from the same lensing survey, with luminous mass values in the same range of the Coma galaxies studied here (see G09a), we show in Fig. \ref{fi02} the physical quantities already plotted there for the Coma galaxies. The two samples reveal almost identical scaling relations between stellar mass-to-light ratio and luminous mass, suggesting a common explanation for the tilt of the FP. The interesting offset between the Coma and SLACS early-type galaxies may be at least in part associated to the different environmental conditions in which the galaxies assembled their stars. In fact, the SLACS galaxies are mainly field galaxies as opposed to the cluster galaxies of Coma. The impact of an older age or a shorter burst of star formation in giving higher values of the mass-to-light ratios for the Coma with respect to the SLACS galaxies is not addressed here and could be checked quantitatively by inspecting the galaxy spectra.

\section{Comments and conclusions}

We have selected a sample of 13 bright Coma galaxies, object of a previous detailed dynamical study to differentiate between the luminous and dark matter components. We have modelled the galaxy SEDs (from the SDSS) with different dust-free solar-metallicity composite stellar population models and derived luminous mass estimates as a function of the assumed IMF. By following the time evolution of the best-fitting SFHs, we have also measured the values of the galaxy luminous mass and $B$-band stellar mass-to-light ratio at redshift $z=0$. We have concluded that:
\begin{enumerate}
\item[$-$] Given the errors, a Salpeter IMF provides luminous mass estimates that are consistent with the dynamical ones, as opposed to the approximately two times smaller values obtained by choosing a ``top-heavy'' IMF such as Kroupa or Chabrier IMFs. 
\item[$-$] The luminous stellar mass-to-light ratio scales with the luminous mass like the corresponding effective quantities are observed to do, thus suggesting that the tilt of the FP can be mainly ascribed to stellar population effects.
\end{enumerate}

A comparison of the photometric and FP analyses performed on this sample of Coma galaxies and different samples of lens galaxies of the SLACS survey (G09a; \citealt{gri08}; \citealt{tre06}) seems to confirm that old, massive early-type galaxies prefer a Salpeter IMF to either a Kroupa or Chabrier IMF and exhibit stellar population properties that are responsible for the the most part for the tilt of the FP. We remark that early-type galaxies that are different in age and mass from the galaxies of our sample may deviate from the picture presented in this analysis.

\section*{Acknowledgements}
We acknowledge the use of data from the SDSS and HyperLeda databases. The SDSS and HyperLeda Web Sites are, respectively, http://www.sdss.org/ and http://leda.univ-lyon1.fr/.

\label{lastpage}

\end{document}